# Connected Health in Multiple Sclerosis: a mobile applications review

Guido Giunti
Salumedia Tecnologias
Seville, Spain
guidogiunti@salumedia.com

Estefania Guisado-Fernandez and Brian Caulfield
University College Dublin
Dublin, Ireland
estefania.guisadofernandez@ucd.ie, b.caulfield@ucd.ie

*Abstract*— **Multiple Sclerosis (MS) is an unpredictable, often disabling disease that can adversely affect any body function; this often requires persons with MS to be active patients who are able to self-manage. There are currently thousands of health applications available but it is unknown how many concern MS. We conducted a systematic review of all MS apps present in the most popular app stores (iTunes and Google Play store) on June 2016 to identify all relevant MS apps. After discarding non-MS related apps and duplicates, only a total of 25 MS apps were identified. App description contents and features were explored to assess target audience, functionalities, and developing entities. The vast majority of apps were focused on disease and treatment information with disease management being a close second. This is the first study that reviews MS apps and it highlights an interesting gap in the current repertoire of MS mHealth resources.**

*mHealth, multiple sclerosis, mobile applications, systematic review; health apps*

I. INTRODUCTION

Multiple sclerosis (MS) is one of the world's most common neurologic disorders, and in many countries it is the leading cause of non-traumatic neurologic disability in young adults[1]. MS is an unpredictable, often disabling disease of the Central Nervous System (CNS) that can adversely affect any body function. However, the most common symptoms are overwhelming fatigue, visual disturbances, altered sensation and difficulties with mobility. Pharmacological treatment of the condition is required but there are many medications and other strategies to manage MS symptoms such as spasticity, pain, bladder problems, fatigue, sexual dysfunction, weakness, and cognitive problems. MS complexity often requires persons with MS to be active patients and able to self-manage[2].

New technologies have indirectly created a new type of patient that many healthcare providers have seen during consults, a patient that is well informed, equipped with electronic communication tools and committed to participate on their own care; an e-patient [3]. This technical growth not only affects patients but the healthcare ecosystem as well. Connected Health is a new model of health management in which, with the support of the new technologies the patient becomes the center of the health care system[4]. The delivery of healthcare or health related services through the use of mobile devices or application, also known as mHealth[5], is included in the Connected Health model. There are currently thousands of mHealth applications[6] available in app stores which has caused researchers to systematically study them (ie Cancer[7], [8]; HIV[9]; Diabetes[10]; etc.). The absence of healthcare professionals involvement in app development continues to be raised time and time again. These concerns revolve around app design and app content alike [11]–[15].

No study has been done for MS apps to this date and the types of mHealth applications available, the intended audience and who are developing apps for this condition are currently unknown. This work describes the current landscape of MS apps and characterizes them based on their features and target audiences.

II. METHODS

*A. Study Design*

A cross-sectional study of MS apps was performed to characterize apps from the two major smartphone app stores: iTunes App Store and Google Play Store, which together represent more than 98.9% of the smartphone app market share[16]. Both stores were systematically searched to identify all relevant apps and provide a systematic presentation and synthesis of the characteristics of the apps. The complete flow of this study is shown in Figure 1.

*B. Setting*

The iTunes App Store serves as the official app store for iOS and has 2 million apps available as of June 2016 [17]. Google Play store (originally the Android Market) serves as the official app store for the Android operating system with over 2.2 million apps available as of June 2016 [17]. On June 17th 2016, iTunes App[18] and Google Play[19] stores from the United States were searched for apps with the keywords "multiple sclerosis" using the audience targeting platform 42matters [20] which aggregates mobile applications data and meta-data across the iOS and Google Play stores.

*C. Selection Criteria*

All apps that were a partial or complete match for the search terms "multiple sclerosis" and the title and/or app store description referred MS or MS related conditions were initially included. Basic and "premium" versions of the same app were considered as separate entries as were versions of the same app for different operating systems. This distinction was considered because of the phenomenon of mobile device fragmentation in which different versions of the same app must co-exist were due to version capabilities or store submission processes. This distinction is also common practice in this type of systematic app reviews[7].

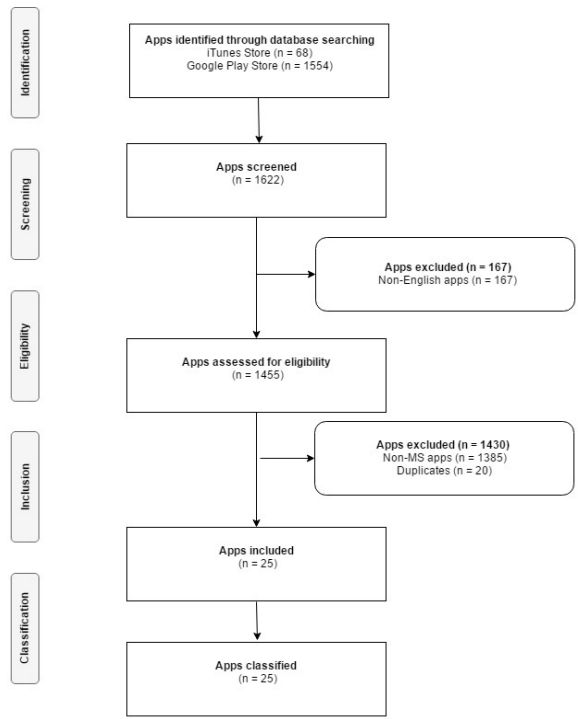

Figure 1. Study Flow.

A small sample (10%) was randomly produced and two reviewers with mHealth experience (GG and EGF) independently evaluated the eligibility of the apps against the selection criteria. In order to assess clarity of the selection criteria, inter-rater reliability was assessed using Fleiss-Cohen's Coefficient. Disagreements were resolved by consensus involving a third reviewer when necessary. After validating the selection criteria and determining inter-rater reliability, the rest of the apps were examined.

GG and EGF independently reviewed the information extracted using structured forms with app title and store descriptions. In cases where these didn't provide enough information, app or developer's websites were visited to extract information on: origin (eg, healthcare related agencies, non-governmental organizations, universities, etc), features and intended audiences.

*1) Inclusion Criteria*
- title and/or description is about MS or MS related conditions

*2) Exclusion Criteria*
- title and/or description is not written in English
- duplicates from the same store

### D. Data Extraction

Data matching the keywords was automatically extracted from the store description of the app using the software application 42matters[20]. Data extracted included app information on: year of release, costs, downloads, ratings, title of app, app description, categories, tags, languages, app websites, screenshots, etc.

### E. Data Coding and Classification

Apps were classified based on their main purpose as described in the store description into only one category following our classification scheme. If the purpose of the app was not clear from the description, a proper classification was discussed among reviewers until consensus was reached.

### F. App Purpose

The app purpose classification scheme is as follows:
- Awareness-raising: tools to raise public recognition of MS as a problem, tools for fundraising, etc.
- Disease and treatment information: provide general information about MS (eg, disease or treatment options)
- Disease management: provide information and practical tools to deal with the medical, behavioral, or emotional aspects of MS.
- Physical Rehabilitation: provide information and practical tools to deal with the medical, behavioral, or emotional aspects of MS.
- Support: provide access to peer or professional assistance.

### G. App Origin

In order to understand how the mHealth developer ecosystem is composed we analyzed and coded the apps based on title, description and developer and/or uploading entity using the following criteria:
- Healthcare related Agency (HCA): hospitals, clinics, pharmaceutical corporations or governmental organizations directly related to healthcare (ie Public Health branches).
- Governmental Agency (GOV): any governmental agency or organization not directly involved in healthcare (ie IT departments).
- Non-governmental Agency (NGO): any organization that is neither a part of a government nor a conventional for-profit business such as societies or organizations that specialize both in general health improvement as well as illness-specific objectives and offer support groups (ie Patient Empowerment Organizations).
- Educational Organizations (EDU): any educational organization such as Universities, Colleges, Libraries or Schools not directly related to healthcare (ie Science School Projects)
- Conferences and Journals (CONF): scientific journals, patient and/or medical conferences.
- Small and Medium-sized Enterprises (SME): startups, software developing companies or any other private organizations that identified themselves as an enterprise and not individuals (ie Digital Health Startups).
- Individuals (IND): developers or uploader entities who are listed as individuals or have not identified themselves as enterprises (ie John Smith).

Whenever discrepancies were found between descriptions and developer or uploading entities, description was considered instead.

### H. App Target Audience

The app descriptions were analyzed to assess intended target audience and coded based on the following criteria:
- Patient-oriented: intended to be used by the general public, patients and/or their family members.
- Clinician-oriented: intended to be used by healthcare professionals or students from health related fields.

### I. Statistical Methods

Categorical variables are presented as absolute and relative frequencies. Continuous variables are presented as mean and standard deviation or median with interquartile range depending on distribution. Landis & Koch's standards for Fleiss-Cohen's Coefficient are used[21]. Statistical analysis was performed using STATA v13.

## III. RESULTS

### A. Selection

A total of 1,622 apps matched the search terms of "multiple sclerosis", of which 68 matches were from the iTunes App Store and 1,554 from the Google Play Store. A random sample (n=168) was independently reviewed by two reviewers (GG and EGF) following the selection criteria. Inter-rater reliability was determined using Cohen's Kappa and found to be more than acceptable at 0.84 (SE 0.03 CI 95% 0.78 – 0.89). One reviewer continued the inclusion/exclusion process with the rest of the apps (n=1,454). After removing duplicates and going through the selection criteria only 25 apps remained (15 for iOS and 10 for Android).

### B. Classification

Table I shows a description of the app population that we explored. The vast majority of MS apps were free (92%) with only 2 iOS apps requiring payment. Android MS apps were rated more often than iOS MS apps.

In terms of MS app functionality, disease and treatment information apps are largely prevalent (close to 70% of all apps); disease management apps are second (almost 25%); awareness raising and support apps are a minority. No physical rehabilitation apps were present. See Table II.

Android MS apps were mostly intended for patients but in iOS apps were more evenly distributed between patient and clinician oriented MS apps. See Figure 2 for some examples of MS apps.

Table III shows the different developing entities by operating system. Small and medium sized enterprises such as start-ups or software companies were responsible for the development of more than half of all MS apps (52%); while a quarter were conferences and publishing agencies (24%). No app was developed by governmental agencies. Table IV shows how MS apps are distributed in terms of their intended audience.

All MS apps found are shown in Table V as is their type of app and app store URL.

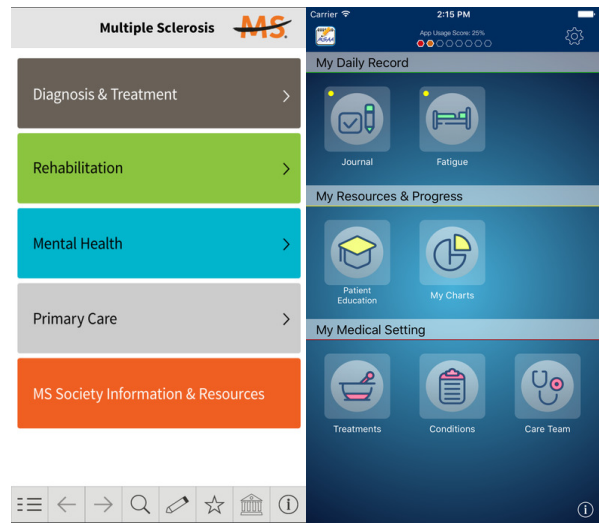

Figure 2. Examples of MS apps. Left: Multiple Sclerosis Diagnosis & Management. Right: MSAA - My MS Manager.

TABLE I. BASIC APP CHARACTERISTICS

|  | *Android* | *iOS* |
|---|---|---|
| **n** | 10 (100%) | 15 (100%) |
| **Free** | 10 (100%) | 13 (86%) |
| **Paid** | - | 2 (14%) |
| **Rated** | 9 (90%) | 5 (33%) |
| **Rating** | | |
| 0 | 1 (10%) | 10 (66%) |
| 1 | - | - |
| 2 | - | 1 (6%) |
| 3 | 1 (10%) | 2 (14%) |
| 4 | 8 (80%) | 2 (14%) |
| 5 | - | - |
| **Number of downloads** | | - |
| 10-50 | 1 (10%) | - |
| 100-500 | 1 (10%) | - |
| 500-1000 | 4 (40%) | - |
| 1000-5000 | 4 (40%) | - |
| Not Available | | 15 |

All percentages are rounded up in accordance with formatting guidelines

TABLE II. APP FUNCTIONALITY BY OPERATING SYSTEM

|  | *Android* | *iOS* |
|---|---|---|
| **Awareness-raising** | 1 (10%) | - |
| **Disease and treatment information** | 6 (60%) | 11 (73%) |
| **Disease management** | 3 (30%) | 3 (20%) |
| **Physical Rehabilitation** | - | - |
| **Support** | - | 1 (7%) |

All percentages are rounded up to follow formatting guidelines

TABLE III. DEVELOPING AGENCY BY OPERATING SYSTEM

|  | **Android** | **iOS** |
|---|---|---|
| **Healthcare related Agency** | 1 (10%) | 1 (7%) |
| **Governmental Agency** | - | - |
| **Non-governmental Agency** | 1 (10%) | - |
| **Educational Organizations** | - | 2 (14%) |
| **Conferences and Journals** | 1 (10%) | 5 (33%) |
| **Small and Medium-sized Enterprises** | 6 (60%) | 7 (46%) |
| **Individuals** | 1 (10%) | - |

All percentages are rounded up to follow formatting guidelines

TABLE IV. APPLICATION TARGET AUDIENCE

|  | **Android** | **iOS** |
|---|---|---|
| **Patient-oriented** | 9 (90%) | 6 (40%) |
| **Clinician-oriented** | 1 (10%) | 9 (60%) |

All percentages are rounded up to follow formatting guidelines

## IV. DISCUSSION

### A. Principal Findings

Persons with MS have a prolonged median survival time from the time of diagnosis of around 40 years[22]. Therefore issues regarding progressive physical and cognitive disability, psychosocial adjustment and social re-integration progress over time. These have implications for persons with MS, their care givers, treating clinicians and society as a whole, in terms of healthcare access, provision of services and financial burden[23], [24]. Many authors have advocated that Smartphone applications have the potential to increase efficiency within medical practice and can provide constantly updated clinical evidence[5], [25]–[28].

To our knowledge, this is the first review of MS mobile applications commercially available to patients, health professionals and public in general. Our study describes the different app functionalities; the proportion of each type; intended audiences; and developing entities.

Considering that MS is one of the world's most common neurologic disorders, how few MS apps are in the app stores (n=25) is in stark contrast with the reality for other conditions such as cancer (n=295 in 2013) [7], diabetes (n=137 in 2009) [29] or HIV (n=124 in 2013) [30] among others.

According to our study, applications for healthcare professionals vary in content, ranging from information for the diagnosis, classification and disease management to educational purposes such as clinical case reports and MRI presentations, conferences, online training and journal apps. On the other hand, patients' apps had a more narrow scope focusing on symptoms information, disease management and only one patient support tool through social networks.

The change in distribution, as seen in Table IV, between patient-oriented and clinician-oriented apps depending on the operating system is intriguing. For other conditions patient-oriented applications usually outnumber those intended for clinicians[7], [31].

The major contributors to the mobile application ecosystem for persons with MS are start-ups and entrepreneurs. It is of note that no governmental agency has launched to market an mHealth solution for MS. Equally interesting is the absence of healthcare provider involvement in app development. Especially considering that the medical community has risen concerns around non-medical personnel promoting app design and development [11]–[15]. It would seem essential that expert medical personnel be involved in the creation of medical apps yet healthcare professionals are seldom involved.

Store's star rating systems and download ranges are frequently used to indicate popularity and indirectly measure the "success" of these apps. However, using those as criteria yields little or no meaningful information on app quality as has been discussed on occasions[32]. The lack of standardized quality measures continues to be concerning, as app use carries risk and can lead to adverse outcomes for both patients and clinicians[33]–[35].

### B. Limitations

One limitation this study has lies in the keyword selection; making it possible for relevant applications to have been unintentionally excluded from our search results. While it's possible that MS might be featured in applications devised for a broader range of conditions (eg. Neurodegenerative Diseases apps), users looking for MS apps would most likely be looking for apps specific to the condition.

The selection of United States app stores only might also have excluded relevant apps. Using Android and iOS based applications exclude the presence of less popular Smartphone platforms like Windows or Blackberry. Google Play and iTunes stores have intrinsic differences that make it difficult to compare certain attributes like number of downloads per app for example. It's also important to note that iTunes App Store and Google Play Store have different processes and steps for app submission which may influence app development. Choosing to focus on only English language applications probably also excluded relevant MS apps from this review.

## V. CONCLUSIONS

There is an interesting gap in the current repertoire of mHealth solutions for persons with MS. It could prove important to address this as it might be a channel to reach persons with MS in a way that's engaging and empowers them. Empowered persons with MS can be more active in their own health management and decision making process.

### ACKNOWLEDGMENT

Guido Giunti and Estefania Guisado-Fernandez gratefully acknowledge the grant number 676201 for the Connected Health Early Stage researcher Support System (CHESS ITN) from the Horizon 2020 Framework Program of the European Commission.

*Pre-print copy accepted at IEEE CMBS – 30th IEEE International Symposium on Computer-Based Medical Systems, 22-24 June 2017, Thessaloniki, Greece – IEEE Copyright*

Authors also would like to thank Luis Fernandez-Luque PhD and Minna Isomursu PhD for their help and collaboration.

TABLE V. APPLICATION TARGET AUDIENCE

| Type | Name of the app | App store URL |
|---|---|---|
| Awareness-raising | Overcoming Multiple Sclerosis | https://play.google.com/store/apps/details?id=com.a11663794335124b9dce47b98a.a92180565a |
| Disease and treatment information | Multiple Sclerosis News | https://play.google.com/store/apps/details?id=com.multiple.sclerosis.news |
| | Multiple Sclerosis Treatment | https://play.google.com/store/apps/details?id=revolxa.inc.multiplesclerosistreatment |
| | Multiple Sclerosis Symptoms | https://play.google.com/store/apps/details?id=com.inc.multiplesclerosissymptoms |
| | Multiple Sclerosis Information | https://play.google.com/store/apps/details?id=multiple.sclerosis.causes.diseases.symptoms.prevention.medicine |
| | Multiple Sclerosis Attack App | https://play.google.com/store/apps/details?id=com.cloudninedevelopmentllc.MSAttack |
| | Multiple Sclerosis Symptoms | https://play.google.com/store/apps/details?id=com.healthappstudio.multiplesclerosissymptoms |
| | Multiple Sclerosis Diagnosis & Management | https://itunes.apple.com/us/app/multiple-sclerosis-diagnosis-management/id480116542?mt=8 |
| | Multiple Sclerosis Attack App | https://itunes.apple.com/us/app/multiple-sclerosis-attack-app/id883546897?mt=8 |
| | eMultipleSclerosis Review | https://itunes.apple.com/us/app/emultiplesclerosis-review/id983062229?mt=8 |
| | Multiple Sclerosis and Related Disorders | https://itunes.apple.com/mx/app/multiple-sclerosis-related/id686600160?mt=8 |
| | MULTIPLE SCLEROSIS - NHS DECISION AID | https://itunes.apple.com/app/multiple-sclerosis-nhs-decision/id585689063?mt=8&ign-mpt=uo%3D4 |
| | Multiple Sclerosis Virtual Education Academy | https://itunes.apple.com/us/app/multiple-sclerosis-virtual-education-academy/id965247008?mt=8 |
| | Multiple Sclerosis Monitor and Commentary | https://itunes.apple.com/ro/app/multiple-sclerosis-monitor-and-commentary/id684850546?mt=8 |
| | Medikidz Explain Multiple Sclerosis | https://itunes.apple.com/us/app/medikidz-explain-multiple-sclerosis/id1011110811?mt=8 |
| | Miniatlas Multiple Sclerosis | https://itunes.apple.com/us/app/miniatlas-multiple-sclerosis/id393496454?mt=8 |
| | Symptoms Of Multiple Sclerosis | https://itunes.apple.com/us/app/symptoms-of-multiple-sclerosis/id1081604312?l=es&mt=8 |
| | 2014 Annual conference in multiple sclerosis | https://itunes.apple.com/do/app/2014-annual-conference-in/id865277898?mt=8 |
| Disease management | MS self Multiple Sclerosis App | https://play.google.com/store/apps/details?id=com.acorda.msself |
| | My Multiple Sclerosis Diary | https://play.google.com/store/apps/details?id=com.appxient.mymsdiary |
| | Multiple Sclerosis EDSS Trackr | https://play.google.com/store/apps/details?id=com.edss |
| | Multiple Sclerosis @Point of Care™ | https://itunes.apple.com/us/app/multiple-sclerosis-point-of-care/id368515953?mt=8 |
| | MS self – Multiple Sclerosis Mobile App for MS Patients | https://itunes.apple.com/us/app/ms-self-multiple-sclerosis-app-for-ms-patients/id744421921?mt=8 |
| | Multiple Sclerosis - MedImage Cases | https://itunes.apple.com/us/app/multiple-sclerosis-medimage-cases/id482713606?mt=8 |
| Support | MyMSTeam: The social network for those who have multiple sclerosis | https://itunes.apple.com/py/app/mymsteam-social-network-for/id628041222?mt=8 |